\documentclass[atoms,article,accept,moreauthors,10pt,a4paper]{mdpi}
\usepackage{booktabs}
\usepackage{multirow}
\usepackage{soul}
\usepackage{microtype}
\usepackage{upgreek}
\setitemize{parsep=6pt,itemsep=0pt,leftmargin=*,labelsep=5.5mm}
\setenumerate{parsep=6pt,itemsep=0pt,leftmargin=*,labelsep=5.5mm}
\setlist[description]{itemsep=0mm}

\usepackage{color}

\newcommand{\cm}{cm$^{-1}$}
\newcommand{\ai}{\textit{ab initio}}

\firstpage{1}
\articlenumber{x}
\doinum{10.3390/------}
\pubvolume{xx}
\pubyear{2018}
\copyrightyear{2018}
\history{Received: 25 February 2018; Accepted: 4 May 2018; Published: date}
 \updates{yes}

\Title{The~ExoMol Atlas of Molecular Opacities}

\Author{Jonathan Tennyson *\orcidA{} and Sergei N. Yurchenko \orcidB{}}
\AuthorNames{Jonathan Tennyson and Sergei N. Yurchenko}

\address[1]{%
Department of Physics and Astronomy,
University College London,
London WC1E 6BT, UK}

\abstract{The~ExoMol project is dedicated to providing molecular line lists for exoplanet and other hot atmospheres.
The~ExoMol procedure uses a mixture of \ai\ calculations and available laboratory data.
The~actual line lists are generated using variational nuclear motion
calculations. These~ line lists form the input for opacity models for cool stars and brown
dwarfs as well as for radiative transport models involving exoplanets.
This~paper is a collection of molecular opacities for  52 molecules (130~isotopologues) at two reference temperatures, 300~K and 2000~K,  using line
lists from the ExoMol database. So far, ExoMol line lists have been generated for about
30 key molecular species. Other  line lists are taken from external sources or
from our work predating the ExoMol project. An overview of the line lists
generated by ExoMol thus far is presented and used to evaluate further molecular
data needs. Other line lists are also considered. The~requirement for
completeness within a line list is
emphasized and needs for further line lists discussed.
}


\begin{document}


\section{Introduction}

Molecular opacities dictate the atmospheric properties and evolution
of a whole range of cool stars and all brown dwarfs. They are also
important for the radiative transport properties of exoplanets. Even
for a relatively simple molecule, the~opacity function is generally
complicated as it varies strongly with both wavelength and temperature.
This~is because molecules interact with light via a variety of
different motions, namely rotational, vibrational and electronic, and
the close-spacing of the energy levels leads to strong temperature
dependence related to the thermal occupancy of the levels.

\textls[-0]{Molecular opacity functions, such as the ones provided by Sharp and
Burrows~\cite{07ShBuxx.dwarfs}, require very extensive input in the
form of temperature-dependent spectroscopic data for atoms and
molecules. At room temperature, the HITRAN database~\cite{jt691}
provides comprehensive and validated spectroscopic parameters for
molecules found in the Earth's atmosphere and some other species.
However, the~complexity of molecular spectra increases rapidly with
temperature and data from HITRAN and can only be used with extreme caution
at higher temperatures~\cite{18ScStHe}. The~HITEMP database~\cite{jt480} is a
HITRAN-style database designed to work at higher temperatures.
However, HITEMP only contains spectroscopic line lists for five
molecules and there are improved line lists available for each of these
species, namely: H$_2$O~\cite{jt734}, CO$_2$~\cite{11TaPe.CO2,17HuScFr.CO2},
CO~\cite{15LiGoRo.CO}, NO~\cite{jt686}, and OH~\cite{16BrBeWe.OH}.
The~opacity of all of these species are discussed further below.}

The~need for extensive, high temperature spectroscopic data on
molecules, many of whom do not occur in the Earth's atmosphere, has
led to a number of systematic efforts to generate the molecular line
lists required
\cite{96JoLaIw.CH,98Plxxxx.TiO,03WeScSt.MgHline,09Bexxxx.exo,jt528,TheoReTS,17HuScFr.CO2}.
In particular, a number of groups have been progressively generating
line lists for key molecules. These~ includes the TheoReTS
\cite{TheoReTS} group (Reims/Tomsk), the~NASA Ames team and our own ExoMol
activity based in University College London, all of which used similar
theoretical procedures discussed below, and an experimental initiative
led by Bernath~\cite{09Bexxxx.exo}. These~ activities have been
significantly enhanced by the discovery of exoplanets and the
requirement of extensive line lists to be used in exoplanet models and
characterization~\cite{jt400,13TiEnCo.exo}.

As molecular line lists are extended, complicated and have a strong
temperature dependence, it can be hard to understand a priori
spectral regions where they show significant absorption. This~work
aims to summarize the absorption by the key molecular species required
for cool stars, brown dwarfs and exoplanets. Many of the line lists  used
are taken from those computed by the Exomol project, but these are
augmented by lists taken from other sources to give a comprehensive
set, or atlas, of absorbing species.

\textls[-15]{In this work, we present a collection (atlas) of molecular opacities for 52 molecules (130~isotopologues) generated using line lists from the ExoMol database. The~opacities are presented in the form of cross sections for two reference temperatures, 300~K and 2000~K using the Doppler (zero-pressure) broadening. The~goal of this atlas is to
present in a simple, visual form an overview of spectroscopic coverage, main spectroscopic signatures as well as temperature dependence of the molecular opacities relevant for atmospheric studies of hot exoplanets and stars.}

\section{Methodology}

The~general methodology used by the ExoMol project is very similar to
that used by both the TheoReTS and NASA Ames group for producing
molecular line lists. It has been well-described elsewhere
\cite{jt511,jt693} and below we will discuss only general
considerations.

Our technique involves the following general steps: (1) computing
accurate \ai\ potential energy and dipole surface moment surfaces, (2)
improving the potential energy surface (PES) using available
high-resolution spectroscopic data and (3) generating a comprehensive
line list using the improved PES, \ai\ dipole moment
surface (DMS) and an appropriate variational nuclear motion program~\cite{jt609,jt338,TROVE}. Development of these program specifically
for opacity calculations has been discussed by us elsewhere~\cite{jt626}.

\textls[-15]{Performing such calculations is always a trade-off between
completeness and accuracy. A~complete} line list should contain
information on possible transitions at all wavelengths and at all
temperatures. An accurate one should reproduce line positions and
transition probabilities to the high standards of laboratory high
resolution spectroscopy. In practice, for most systems, it is hard if not
impossible to achieve both of these goals simultaneously so
compromises have to be made. Experience shows that completeness is
essential~\cite{jt572} while accuracy is something to strive for and,
of course, is essential for high resolution spectroscopic studies
\cite{15HoDeSn.TiO}.

Starting from an \ai\ PES, there are three ways of improving
the accuracy of the line positions. The~first method is to fit the
potential to measured transitions frequencies, empirical energies or a
mixture of both. This~is now a standard technique that is widely
employed. It is capable of giving very good results~\cite{ps97} if in
most cases not actual experimental accuracy. The~second technique is
to adjust the vibrational band origins during the calculation. This
technique works only if the rotation--vibration basis used is a simple
product between vibrational and rotational functions. A basis in this
form is used by the polyatomic nuclear motion program TROVE
\cite{TROVE} and the vibrational basis is contracted by performing an
initial rotationless ($J=0$) calculations. At this stage, the band
origins can be shifted to their empirical value~\cite{jt503}. Finally,
the ExoMol data structure~\cite{jt548} presents a line list as a states
file with energy levels and associated quantum numbers and a highly
compact transitions file of Einstein A coefficients. This~structure
therefore allows computed energy levels to be replaced with empirical
ones after the calculation is complete; indeed, this can be done some
time after the line list is computed if new empirical data becomes
available~\cite{jt570}. This~allows many transition frequencies to be
generated with experimental accuracy, including ones that have not
actually been measured.

To provide lists of accurate empirical energy levels, Furtenbacher~et~al.
\cite{jt412,12FuCsa} developed the measured active rotation vibration
energy level procedure (MARVEL). Table~\ref{tab:marvel}
gives a summary of molecules of astronomical importance for which the
MARVEL procedure has been applied. Overviews of this methodology
can be found elsewhere~\cite{07CsCzFu.marvel,jt562}.

\begin{table}[H]
\caption{Summary of MARVEL analyses performed for astronomically important molecules.}
\label{tab:marvel}
\centering
\begin{tabular}{ccrrc}
\toprule
\bf Molecule&\boldmath{$N_{\rm iso}$}&\boldmath{$N_{\rm trans}$}&\boldmath{$N_{E}$}&\bf Reference(s)\\
\midrule
H$_2$O & 9 & 182,156 & 18,486 &~\cite{jt454,jt482,jt539,jt576}\\
H$_3^+$& 3 & 1410 & 911& ~\cite{13FuSzMa.marvel,13FuSzFa.marvel}\\
NH$_3$ & 1 & 28,530 & 4961&~\cite{jt608}\\
C$_2$ & 1 & 23,343 & 5699 &~\cite{jt637}\\
TiO & 1 & 48,590& 10,564&~\cite{jt672}\\
HCCH & 1 & 37,813&11,213&~\cite{jt705}\\
SO$_2$& 3 &40,269& 15,130&~\cite{jt704}\\
H$_2$S& 1 &39,267&7651&~\cite{jt718}\\
ZrO&1& 21,195& 8329 &~\cite{jtZrO}\\
\bottomrule
\end{tabular}

\begin{tabular}{@{}c@{}}
\multicolumn{1}{p{\textwidth -.88in}}{\footnotesize $N_{\rm iso}$: Number of~isotopologues considered; $N_{\rm trans}$: Number of transitions validated for the main isotopologue;
$N_E$:  Number of energy levels given for the main isotopologue.}
\end{tabular}

\end{table}

The~line lists generated by the methods discussed above form the input
to opacity calculations. To~judge the potential influence of each molecule
on  the opacity,
one can generate cross sections as a function of wavelength and
temperature. As several of the line lists considered contain many billions
of lines (see Table~\ref{tab:exomoldata}), calculating cross sections can be
computationally expensive. For this reason, we have developed a highly optimized
program  ExoCross~\cite{jt708}, which generates wavelength-dependent
cross sections as function of temperature and pressure. Here, we use
ExoCross to systematically generate cross sections for the key species
that are important for opacities of cool stars, brown dwarfs and~exoplanets.

In generating these cross sections, we consider the main (parent) isotopologue
of each species, which is taken as being in 100 \%\ abundance. For simplicity, we use a Doppler profile on a wavenumber grid of 1~\cm. The~cross sections have been generated using the methodology by Hill~et~al. ~\cite{jt542}. The~cross sections can be also obtained at higher resolutions (up to 0.01~\cm) using the cross sections App at \href{www.exomol.com}{www.exomol.com}.

Cross sections were generated at two standard temperatures of 300 K and 2000 K
except when the molecule is  expected to be entirely in the condensed
phase at 300 K. For a few, key strongly bound species, we also consider
the cross section at 5000 K, which is near the upper limit of the temperature
where molecules make a contribution to opacities.

\section{Results}

Table~\ref{tab:exomoldata} summarises the molecules for which line lists
have been provided as part of the ExoMol project. Sources for line lists
of other key species are given in Table~\ref{tab:otherdata}.
Data for all these species, including cross sections, are
available on the ExoMol website.

\begin{figure}[H]
\centering
\includegraphics[width=0.99\columnwidth]{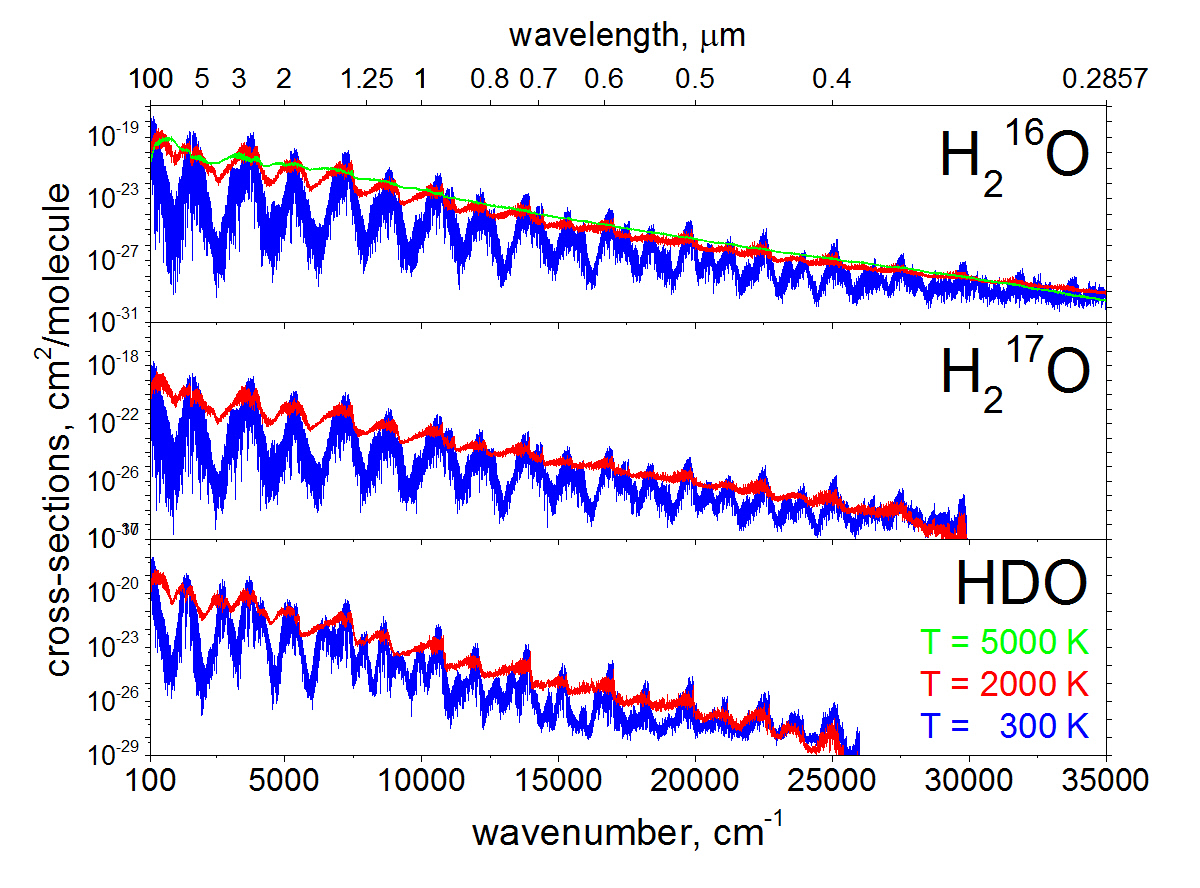}
\caption{Cross sections for H$_2$$^{16}$O from the POKAZATEL line list
 ~\cite{jt734}, H$_2$$^{17}$O from the HotWat78 line list~\cite{jt665}
  and HDO from the VTT line list~\cite{jt469}. All cross sections are
  for 100\%\ abundance.}
\label{f:water}
\end{figure}

\begin{figure}[H]
\centering
\includegraphics[width=0.93\columnwidth]{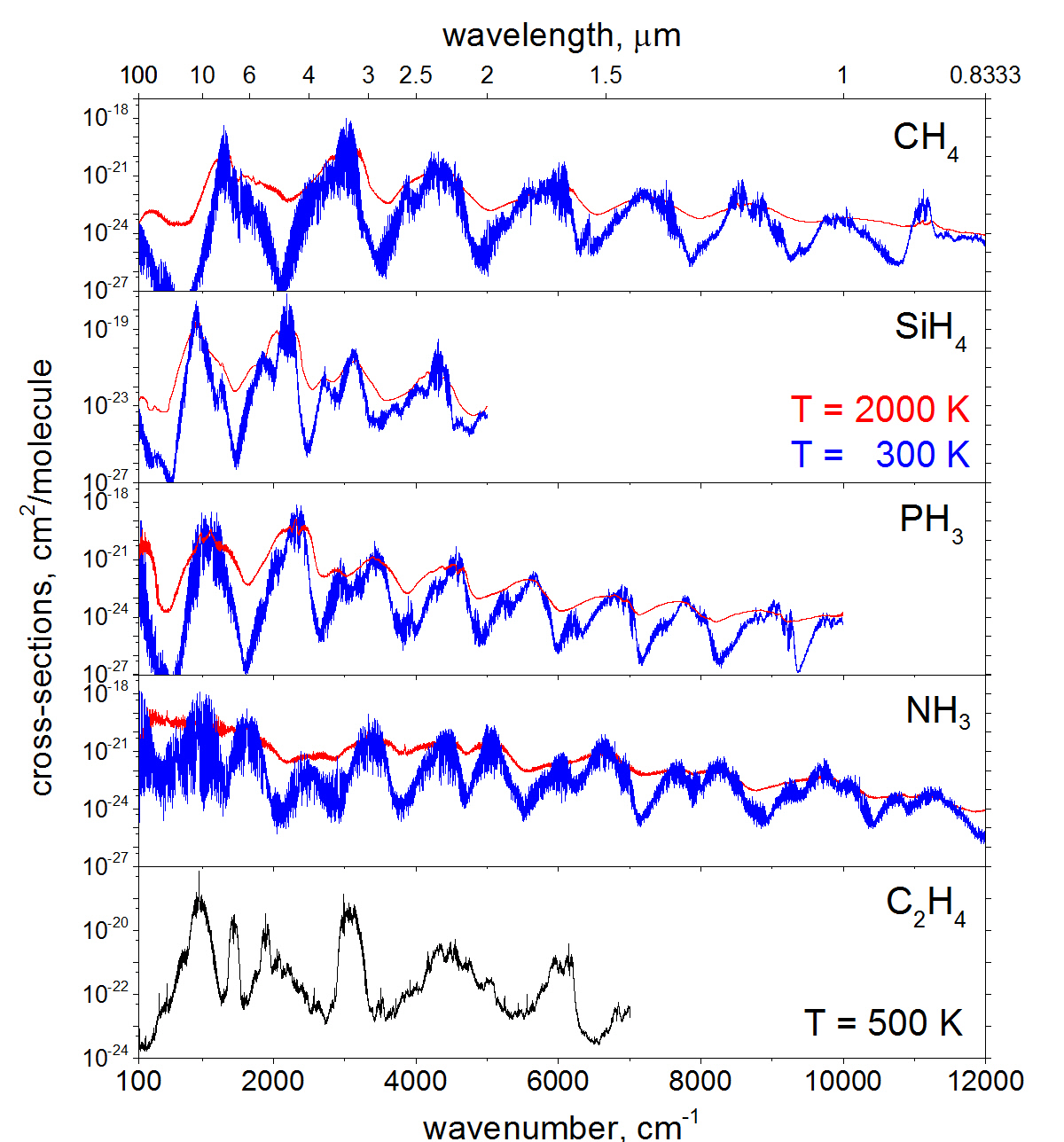}
\caption{Cross sections generated using ExoMol line lists
for methane, 10to10 line list ~\cite{jt564}, silane
\cite{jt711}, phosphine~\cite{jt592}, ammonia~\cite{jt500} and ethylene~\cite{jt729}.}
\label{f:polyhydrides}
\end{figure}

\begin{figure}[H]
\centering
\includegraphics[width=0.98\columnwidth]{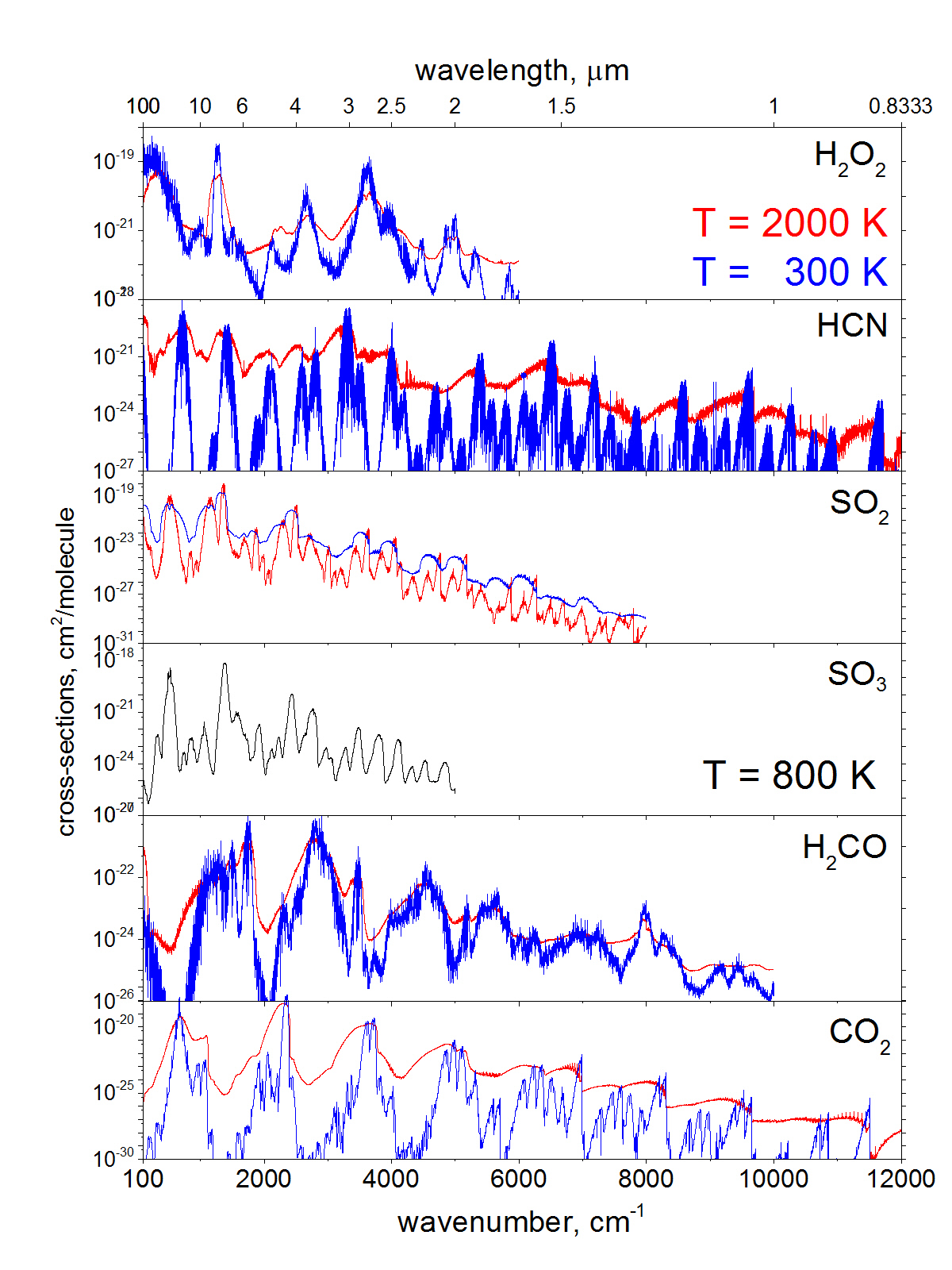}
\caption{Cross sections for polyatomic oxides and HCN. Line lists
are from ExoMol for hydrogen peroxide~\cite{jt620}, hydrogen
cyanide~\cite{jt570}, sulfur dioxide~\cite{jt635}, nitric aid~\cite{jt614}  and sulfur
trioxide~\cite{jt641}. The~carbon dioxide data is taken from Ames-2016~\cite{17HuScFr.CO2}.}
\label{f:polyoxides}
\end{figure}

\begin{figure}[H]
\centering
\includegraphics[width=0.98\columnwidth]{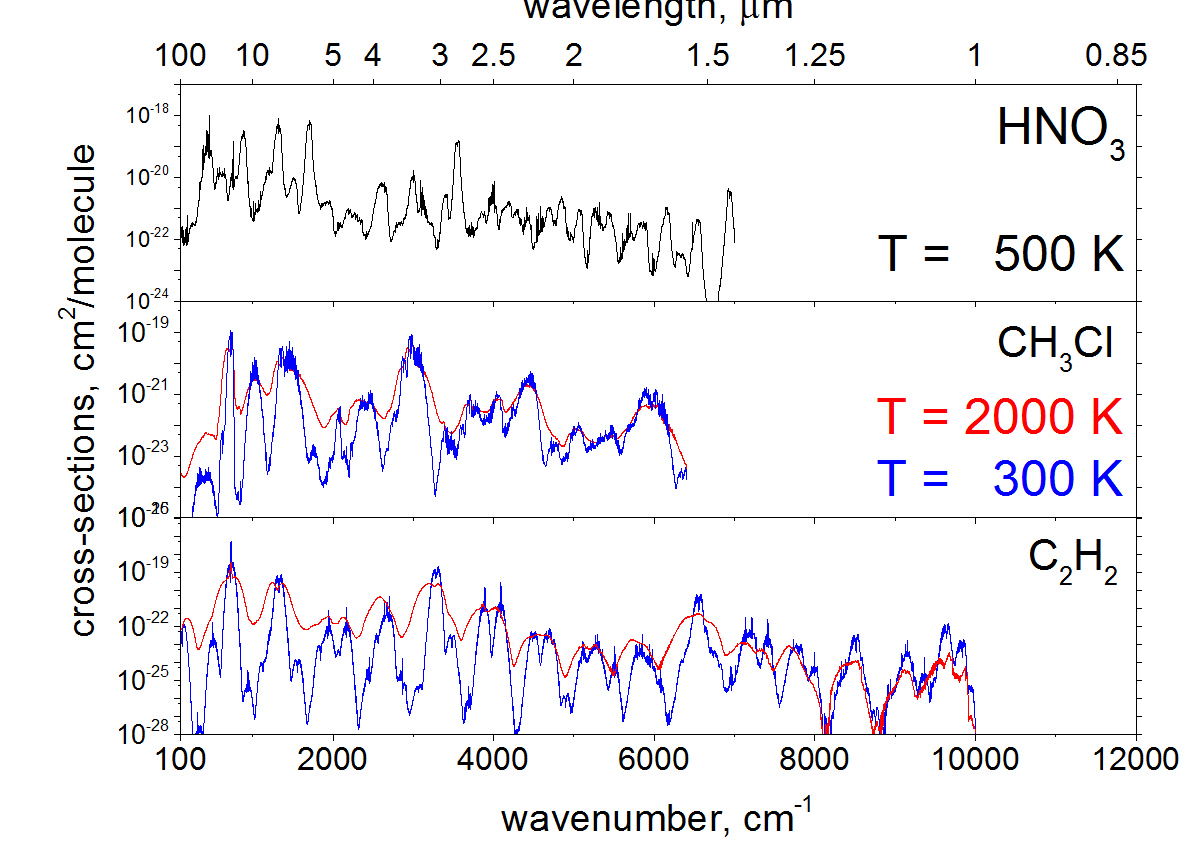}
\caption{Cross sections obtained from ExoMol line lists for
 HNO$_3$~\cite{jt614},
CH$_3$Cl~\cite{jt729}, and C$_2$H$_2$~\cite{17LyPe.C2H2}.}
\label{f:HNO3:CH3Cl:C2H2}
\end{figure}

\begin{figure}[H]
\centering
\includegraphics[width=0.98\columnwidth]{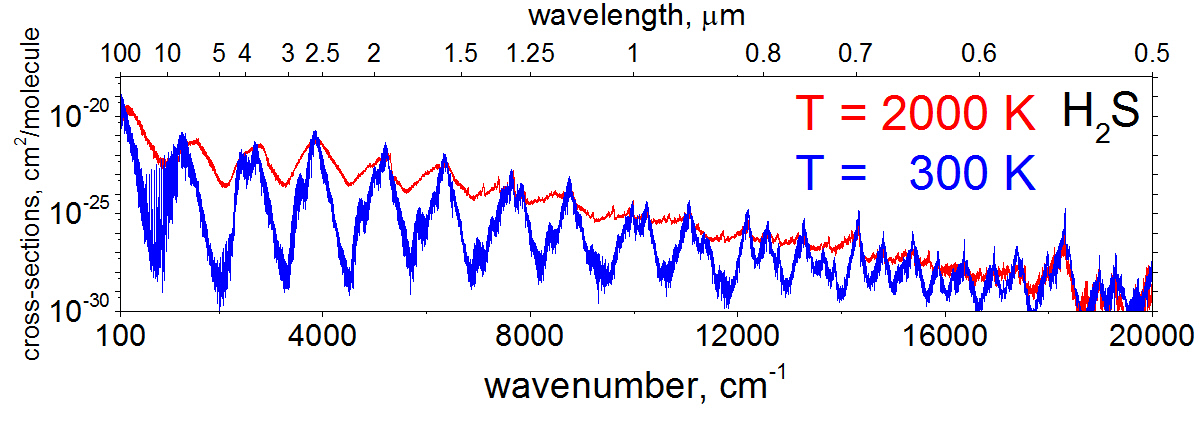}
\caption{Cross sections for hydrogen sulfide
generated using the ExoMol line list
for H$_2$S~\cite{jt640}. }
\label{f:H2S}
\end{figure}

\begin{figure}[H]
\centering
\includegraphics[width=0.98\columnwidth]{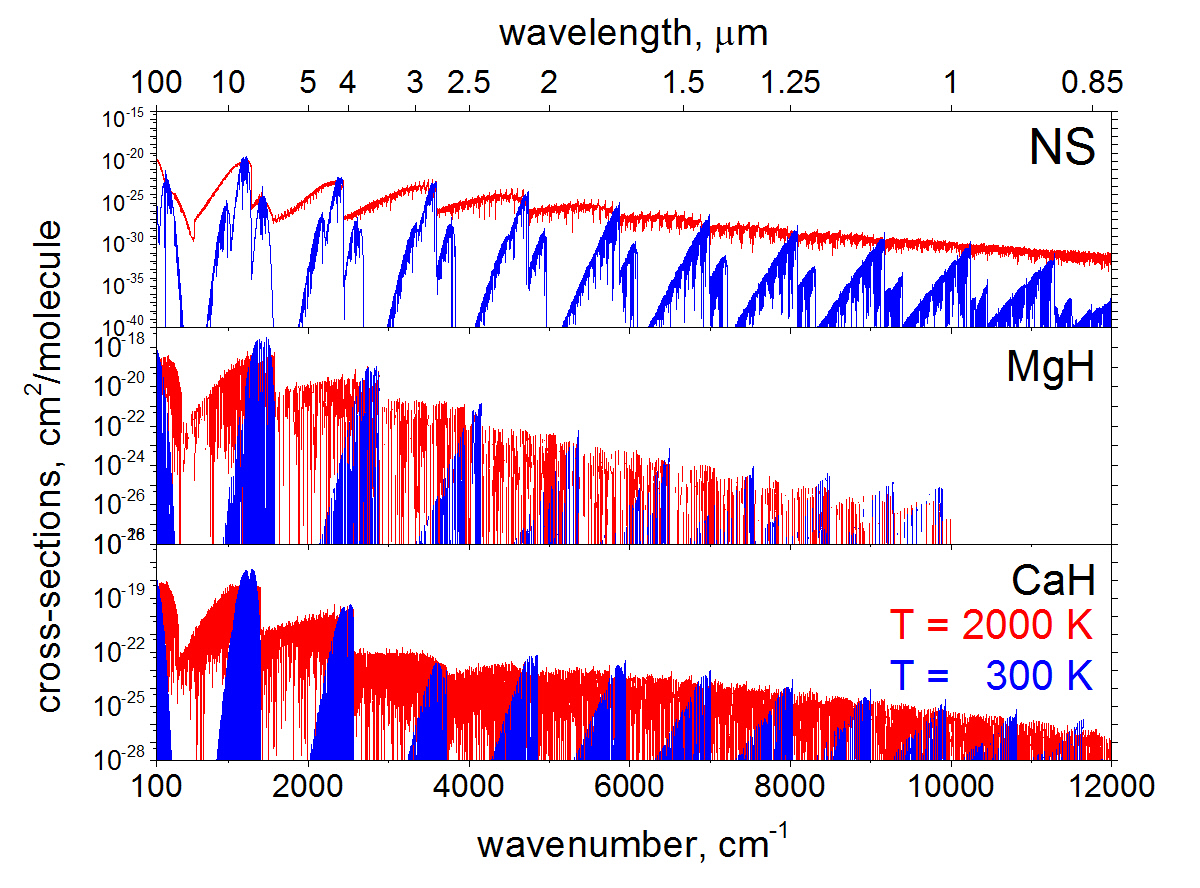}
\caption{Cross sections for alkaline earth monohydrides MgH and
CaH from the Yadin ExoMol line lists~\cite{jt529} and NS from the SNaSH line list~\cite{jt725}.}
\label{f:XH}
\end{figure}

\begin{figure}[H]
\centering
\includegraphics[width=0.98\columnwidth]{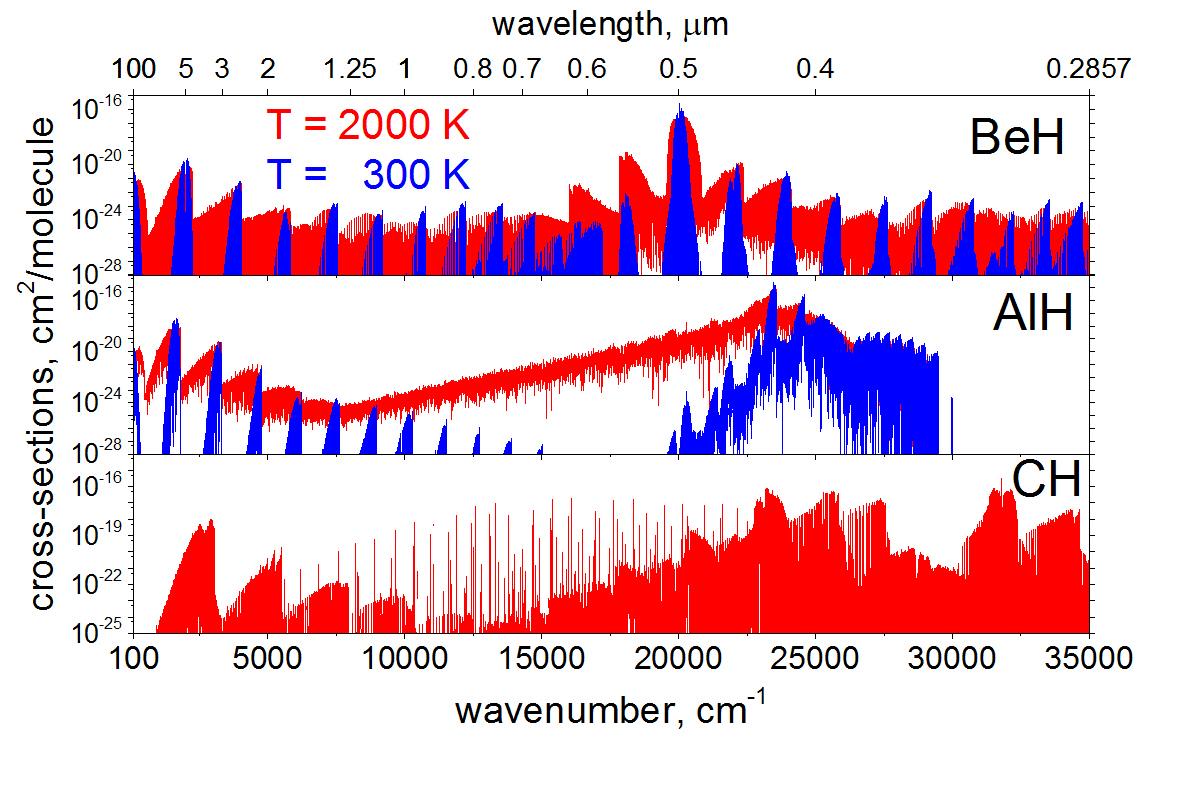}
\caption{Cross sections for alkaline earth monohydrides and CH.
 BeH uses the updated
ExoMol line list of  Darby-Lewis~et~al.~\cite{jt722}, AlH is the new ExoMol line list~\cite{jt732} and CH is the empirical work of  Masseron~et~al.~\cite{14MaPlVa.CH}; the CH line list  is
only defined for $T>1000$~K.}
\label{f:BeH:AlH:CH}
\end{figure}

\begin{figure}[H]
\centering
\includegraphics[width=0.98\columnwidth]{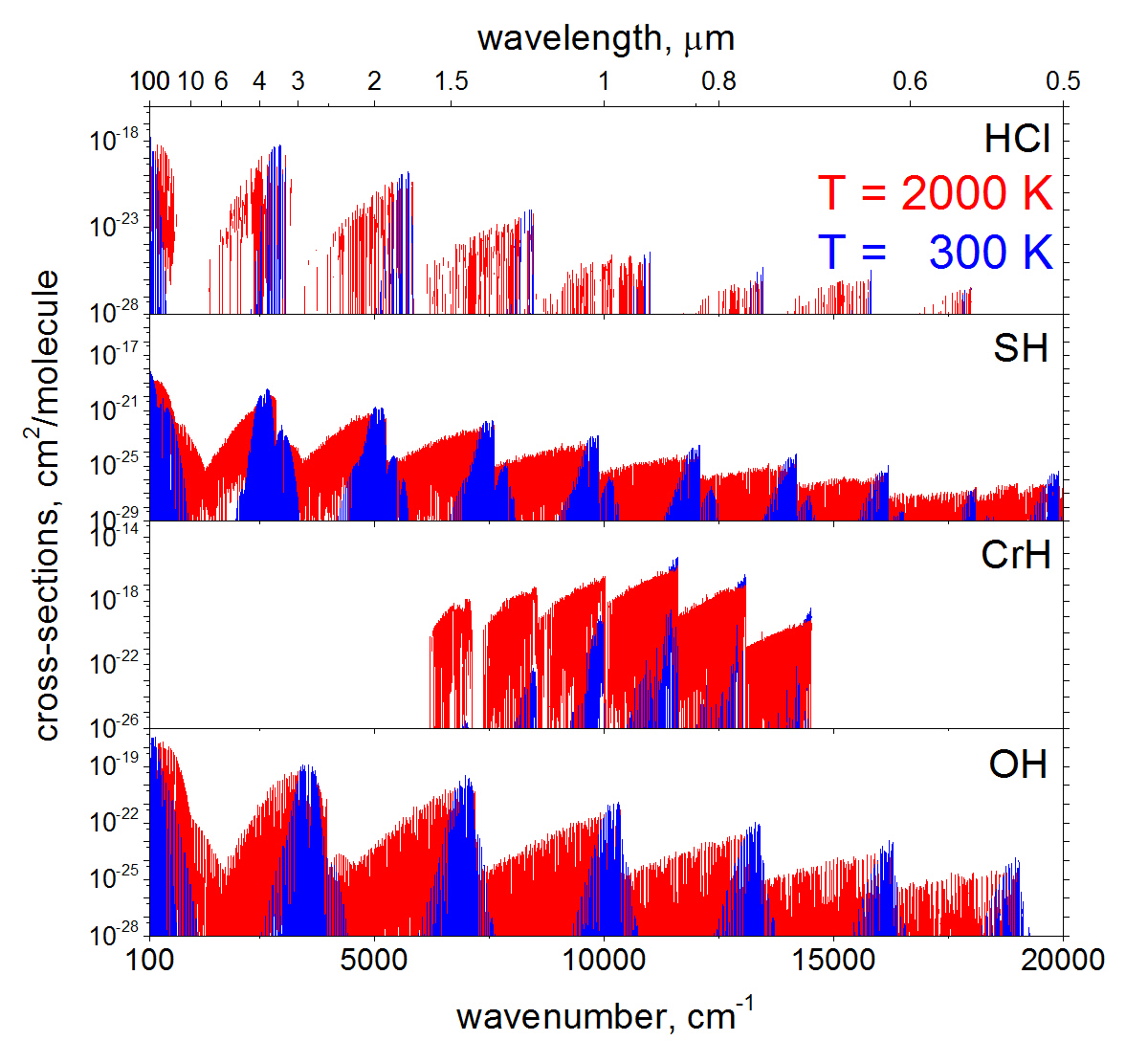}
\caption{\textls[-15]{Cross sections for  monohydrides: an empirical list due to Li.~et~al.
\cite{13LiGoHa.HCl} for HCl, an ExoMol line list for mercapto radical SH~\cite{jt725},
chromium hydride~\cite{02BuRaBe.CrH}. The OH data are taken from HITEMP~\cite{jt480}}.}
\label{f:HCl:HS:CrH:OH}
\end{figure}

\begin{figure}[H]
\centering
\includegraphics[width=0.98\columnwidth]{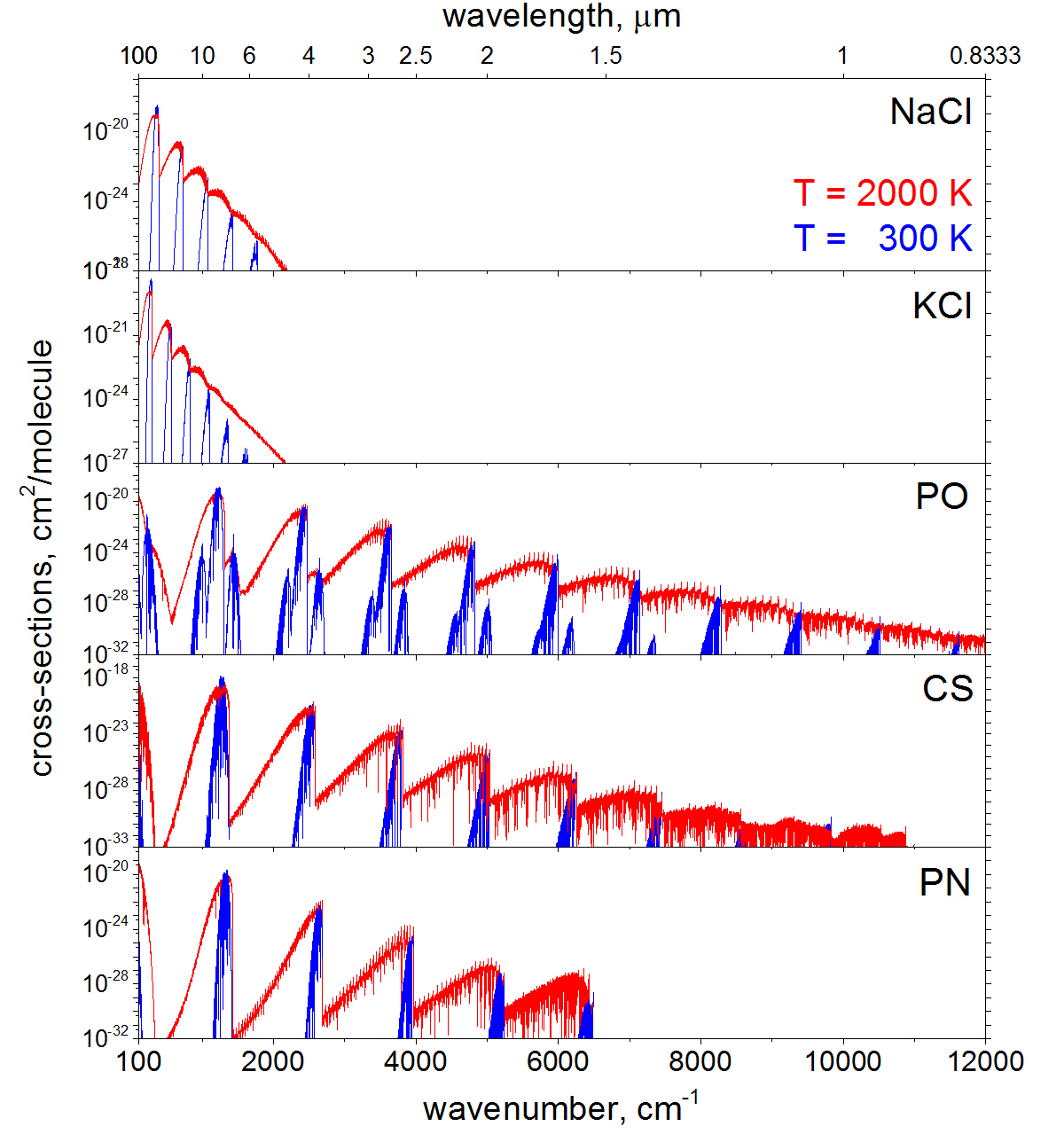}
\caption{Cross sections generated from ExoMol line lists for
sodium chloride~\cite{jt583}, potassium chloride~\cite{jt583}, phosphorous
monoxide ~\cite{jt703}, carbon monosulfide~\cite{jt615} and phosphorous nitride~\cite{jt590}.}
\label{f:polar}
\end{figure}

\begin{figure}[H]
\centering
\includegraphics[width=0.98\columnwidth]{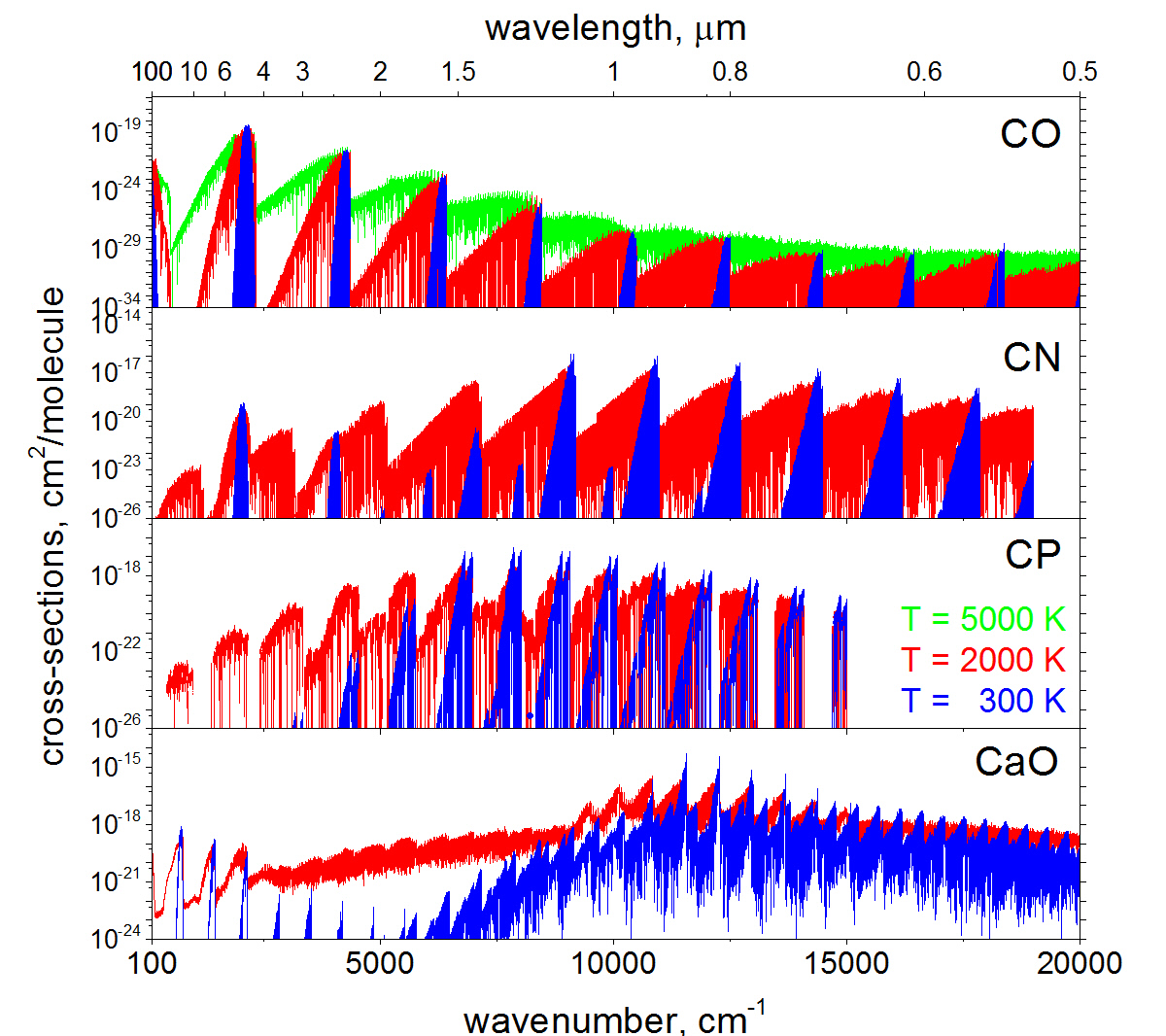}
\caption{Cross sections for carbon monoxide, cyanide, carbon phosphide and
calcium oxide. The~CN~\cite{14BrRaWe.CN} and
CP~\cite{14RaBrWe.CP} cross sections are based on empirical line lists
from the Bernath group. The~CaO data are taken from an ExoMol line list
\cite{jt618}. The CO~\cite{15LiGoRo.CO} line list is based on an empirical dipole moment~function.}
\label{f:cdiatomics}
\end{figure}

\begin{figure}[H]
\centering
\includegraphics[width=.99\columnwidth]{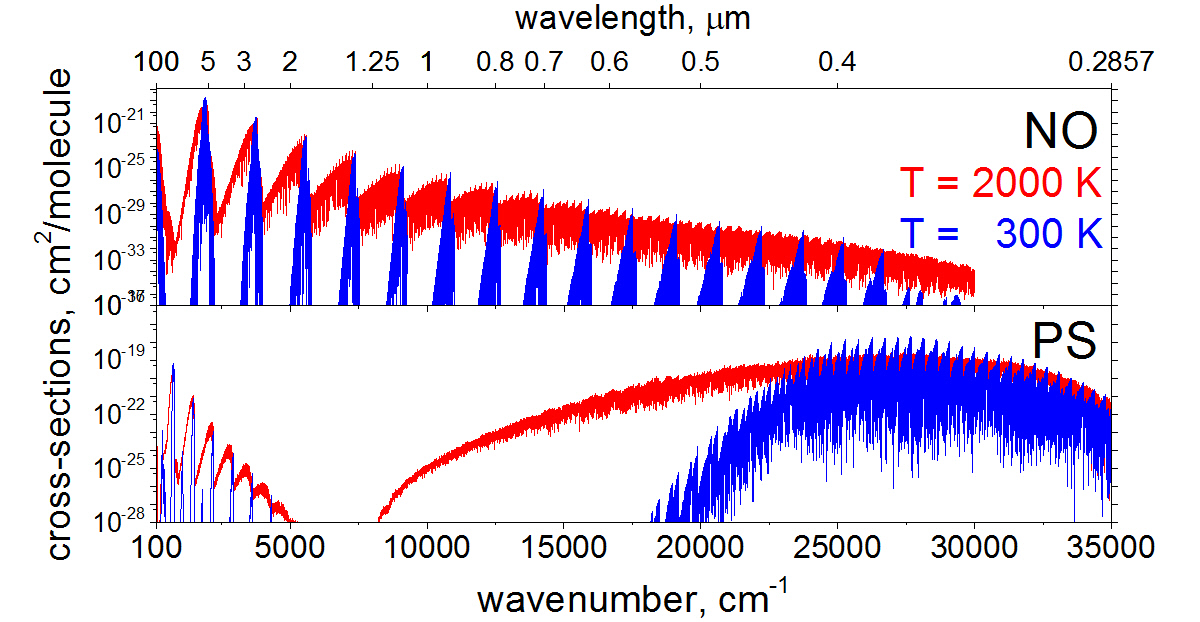}
\caption{Cross sections generated from ExoMol line lists for nitric oxide
\cite{jt686} and phosphorous monosulfide~\cite{jt703}.}
\label{f:NO}
\end{figure}

\begin{figure}[H]
\centering
\includegraphics[width=0.98\columnwidth]{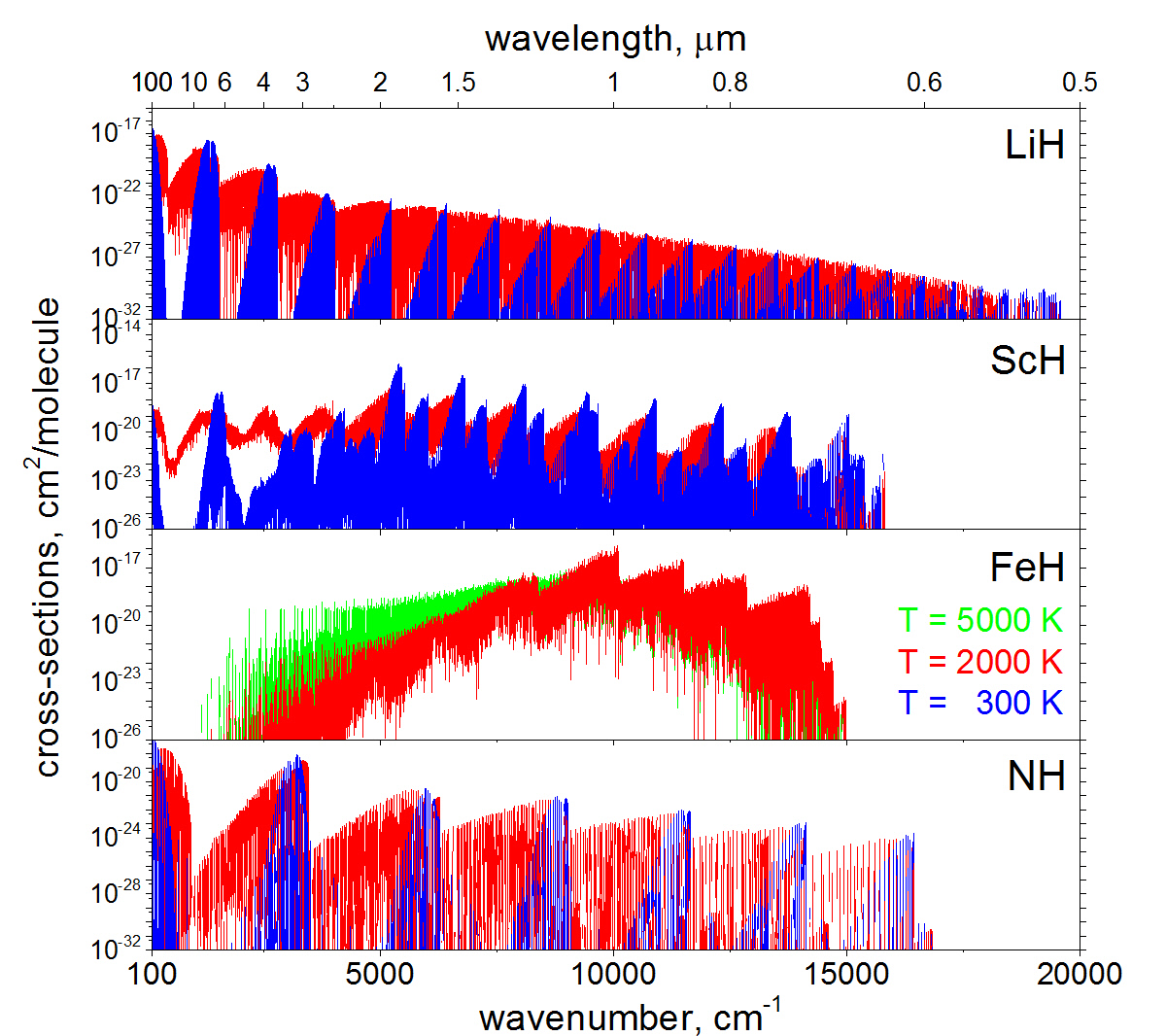}
\caption{Cross sections for metal hydrides and NH. Line lists
for  lithium hydride~\cite{jt506} and scandium hydride~\cite{jt599}
are theoretical while those for FeH and NH are derived from the experiments of
the Bernath group~\cite{03DuBaBu.FeH,10HaHiBa.FeH,10WEReSe.FeH,14BrBeWe.NH}.}
\label{f:ZH}
\end{figure}

\begin{figure}[H]
\centering
\includegraphics[width=0.98\columnwidth]{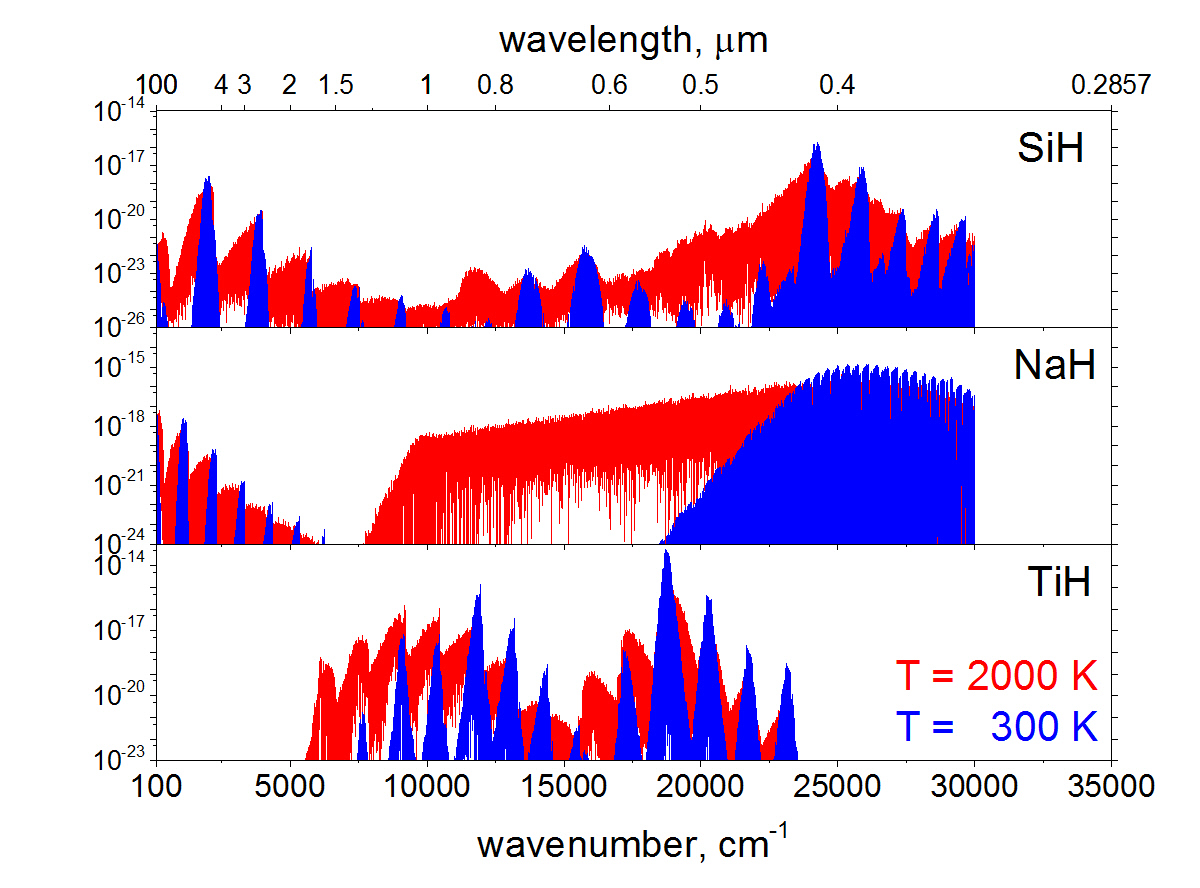}
\caption{Cross sections for hydride species
based on ExoMol line lists for sodium hydride
\cite{jt605} and silicon monohydride~\cite{jt711}, and the empirical
titanium monohydride line list of \mbox{Burrows~et~al.}~\cite{05BuDuBa.TiH}.}
\label{f:YH}
\end{figure}

\begin{figure}[H]
\centering
\includegraphics[width=0.98\columnwidth]{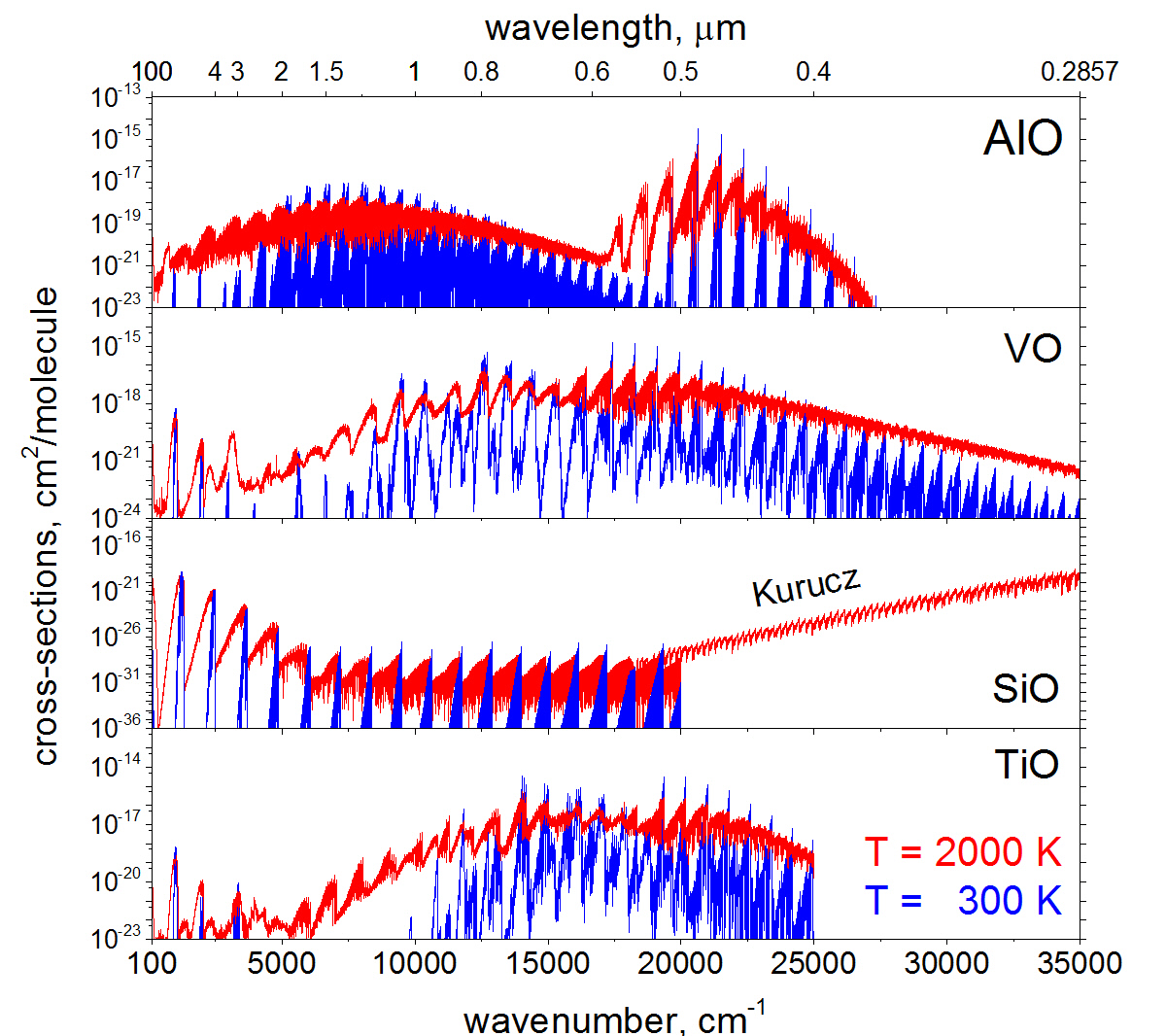}
\caption{Cross sections for metal oxides generated using ExoMol
line lists for silicon monoxide~\cite{jt563}, aluminium monoxide ~\cite{jt598} and vanadium monoxide~\cite{jt644}. The~titanium monoxide cross
sections are based on the computed line list due to Schwenke~\cite{98Scxxxx.TiO}. Also shown are short-wavelength silicon monoxide cross sections
generated using line data from the database due to Kurucz~\cite{11Kurucz.db}.}
\label{f:XO}
\end{figure}

\begin{figure}[H]
\centering
\includegraphics[width=.98\columnwidth]{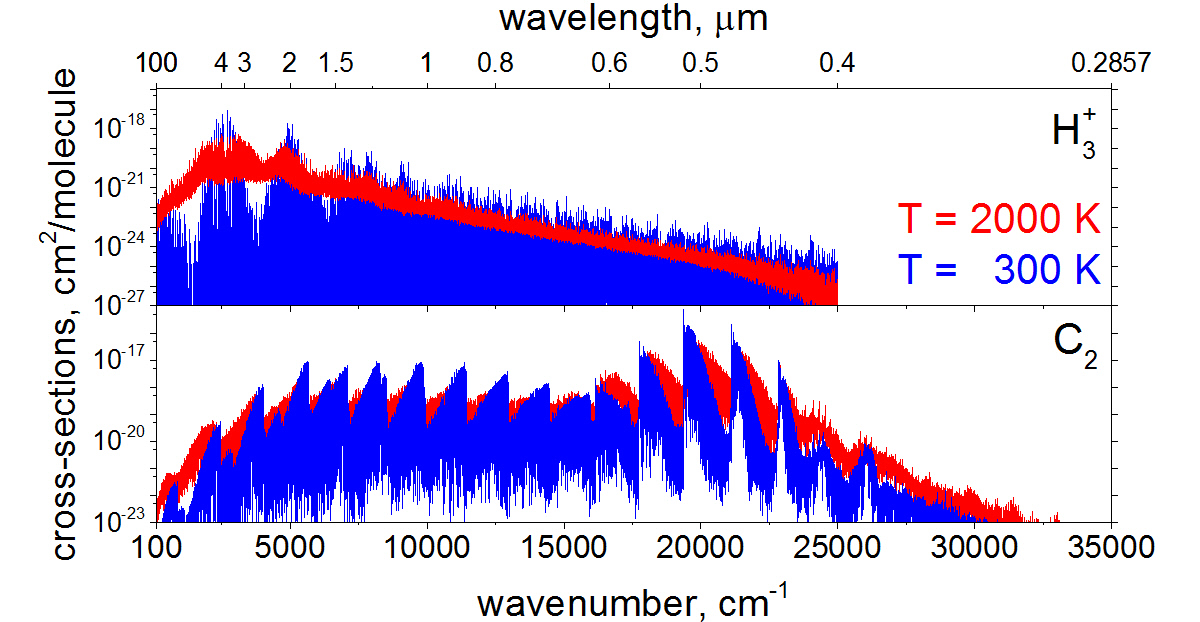}
\caption{Cross sections based on ExoMol line lists for carbon dimer~\cite{jtexoC2} and H$_3^+$ molecular ion ~\cite{jt666}. }
\label{f:uv}
\end{figure}

Figures~\ref{f:water}--\ref{f:uv} display temperature-dependent cross
sections for each of these species. For all molecules except water, only results for the major
(parent) isotopologue are given. As specified in  Table~\ref{tab:exomoldata},
many ExoMol line lists also consider isotopically substituted species.
In most cases, isotopic substitution leads to shifts in spectroscopic
features that are observable at medium to high resolution but result in no
fundamental change in structure of the spectrum.
The~exception is where this substitution
leads to breaking of the symmetry, which can lead to new features
due to changes in the way the Pauli principle applies, examples here
include $^{12}$C$^{12}$C to $^{12}$C$^{13}$C, and changes in vibrational
structure such as those encountered on moving from H$_2$O to HDO.

\begin{table}[H]
\caption{Datasets created by the ExoMol project and included in the ExoMol database~\cite{jt631}.}
\label{tab:exomoldata}
\centering
\begin{tabular}{llrrrrlc}
\toprule
\bf Paper&\bf Molecule&\boldmath{$N_{\rm iso}$}&\boldmath{$T_{\rm max}$}&\boldmath{$N_{\rm elec}$}&\boldmath{$N_{\rm lines}$}
&\bf DSName&\bf Reference\\
\midrule
I&BeH&1&2000 &1&16,400&Yadin& ~\cite{jt529}\\
I&MgH&3 &2000 &1&10,354&Yadin& ~\cite{jt529}\\
I&CaH&1 &2000 &1&15,278&Yadin& ~\cite{jt529}\\
II&SiO&5&9000&1& 254,675&EJBT&~\cite{jt563}\\
III&HCN/HNC&1&4000&1&399,000,000&Harris& ~\cite{jt570}\\
IV&CH$_4$&1&1500&1&9,819,605,160&YT10to10& ~\cite{jt564}\\
V&NaCl&2&3000&1& 702,271 &Barton&~\cite{jt583}\\
V&KCl&4&3000&1& 1,326,765  &Barton&~\cite{jt583}\\
VI&PN&2&5000&1&142,512&YYLT&~\cite{jt590}\\
VII&PH$_3$&1&1500&1&16,803,703,395&SAlTY&~\cite{jt592}\\
VIII&H$_2$CO&1&1500&1&10,000,000,000&AYTY&~\cite{jt597}\\
IX&AlO&4&8000&3&4,945,580&ATP& ~\cite{jt598}\\
X&NaH&2&7000&2&79,898&Rivlin&~\cite{jt605}\\
XI&HNO$_3$&1&500&1&6,722,136,109&AlJS&~\cite{jt614}\\
XII&CS&8&3000&1&548,312&JnK&~\cite{jt615}\\
XIII&CaO&1&5000&5&21,279,299&VBATHY&~\cite{jt618}\\
XIV&SO$_2$&1&2000&1&1,300,000,000&ExoAmes& ~\cite{jt635}\\
XV&H$_2$O$_2$&1&1250&1&20,000,000,000&APTY& ~\cite{jt638}\\
XIV&H$_2$S&1&2000&1&115,530,3730& AYT2& ~\cite{jt640}\\
XV&SO$_3$&1&800&1&21,000,000,000&UYT2& ~\cite{jt641}\\
XVI&VO&1&2000&13&277,131,624&VOMYT& ~\cite{jt644}\\
XIX&H$_2$$^{17,18}$O&2&3000&1&519,461,789&HotWat78& ~\cite{jt665}\\
XX&H$_3^+$&1&3000&1&11,500,000,000&MiZATeP& ~\cite{jt666}\\
XXI&NO&6&5000&2&2,281,042&NOName& ~\cite{jt686}\\
XXII&SiH$_4$&1&1200&1&62,690,449,078&OY2T& ~\cite{jt701}\\
XXIII&PO&1&5000&1&2,096,289&POPS&~\cite{jt703}\\
XXIII&PS&1&5000&3&30,394,544&POPS& ~\cite{jt703}\\
XXIV&SiH&4&5000&3&1,724,841&SiGHTLY& ~\cite{jt711}\\
XXV&SiS&12&5000&1&91,715&UCTY&\cite{jt724}\\
XXVI&HS&6&5000&1&219,463&SNaSH&\cite{jt725}\\
XXVI&NS&6&5000&1&3,479,067&SNaSH&\cite{jt725}\\
XXVII&C$_2$H$_4$&1&700&1&49,841,085,051&MaYTY&\cite{jt729}\\
XXVIII&AlH&3&5000&3&40,000&AlHambra&\cite{jt732}\\
XXIX&CH$_3$Cl&2&1200&1&166,279,593,333& OYT &~\cite{jt729} \\
XXX&H$_2$$^{16}$O&1&5000&1&1,500,000,000&POPKAZATEL& ~\cite{jt734}\\
XXXI&C$_2$&3&5000&8&6,080,920&8State&~\cite{jtexoC2}\\
XXXII&MgO&3&5000&4&22,579,054&LiPTY&~\cite{jtMgO}\\
\bottomrule
\end{tabular}

\begin{tabular}{@{}c@{}}
\multicolumn{1}{p{\textwidth -.88in}}{\footnotesize $N_{\rm iso}$: Number of~isotopologues considered;
$T_{\rm max}$: Maximum temperature for which the line list is complete;
$N_{\rm elec}$: Number of electronic states considered;
$N_{\rm lines}$:  Number of lines, value is for the main isotopologue;
DSName: Name of line list and of data set in ExoMol database~\cite{jt631}.}
\end{tabular}
\end{table}

For most species, absorption cross sections are plotted at two temperatures: 300 K and 2000 K.
Exceptions are where the species is unlikely to have significant vapor pressure at 300 K, when
this curve has been omitted, or when the species unlikely to survive at 2000 K, in which case
an appropriate, lower temperature is used.
The~figures are grouped according to whether the line lists concerned cover the
infrared only (wavenumbers below 12,000 \cm), extend into the visible (wavenumbers below
20,000 \cm) or cover the near ultraviolet (wavenumbers below 35,000 \cm). Beyond
that, the figures are grouped to contain roughly similar species.
Each of the figures and species are considered in turn~below.

\begin{table}[H]
\centering
\captionsetup{width=1\linewidth}

\caption{Other molecular line lists considered. These data can also be obtained from the ExoMol website. }
\label{tab:otherdata}
\scalebox{.95}[.95]{\begin{tabular}{lrrrrcccc}
\toprule

\multirow{2}{*}{\bf Molecule\vspace{-6pt}}
 &\multirow{2}{*}{\boldmath{$N_{\rm iso}$}\vspace{-6pt}}&\multirow{2}{*}{\boldmath{$T_{\rm max}$}\vspace{-6pt}}&\multirow{2}{*}{\boldmath{$N_{elec}$}\vspace{-6pt}}&\multirow{2}{*}{\boldmath{$N_{\rm
lines}$}\vspace{-6pt}}&\multirow{2}{*}{\bf DSName\vspace{-6pt}}&\multirow{2}{*}{\bf Reference\vspace{-6pt}}& \multicolumn{2}{c}{\bf Methodology}\\\cmidrule(lr){8-9}
&&&&&&& \bf Line Positions & \bf Intensities \\
\midrule

NH$_3$&2&1500&1&1,138,323,351&BYTe&~\cite{jt500}& empirical $^a$ & \ai\,$^a$ \\
LiH&1&12,000&1&18,982&CLT&\cite{jt506}&\ai & \ai \\
ScH&1&5000&6&1,152,827&LYT&\cite{jt599}&tuned \ai & \ai\\
NH&1&&1&10,414&14BrBeWe&\cite{14BrBeWe.NH}&empirical & \ai \\
CH&2&6000&4&54,086&14MaPlVa&\cite{14MaPlVa.CH}&empirical & \ai \\
CO&9&9000&1&752,976&15LiGoRo&\cite{15LiGoRo.CO}&empirical & emp./\ai\,$^b$\\
OH&1&6000&1&45,000&16BrBeWe&~\cite{16BrBeWe.OH}&empirical & \ai\\
CN&1&&1&195,120&14BrRaWe&\cite{14BrRaWe.CN}&empirical & \ai\\
CP&1&&1&28,735&14RaBrWe&\cite{14RaBrWe.CP}&empirical & \ai\\
HCl&1&&1&2588&11LiGoBe&\cite{13LiGoHa.HCl}&empirical & \ai\\
FeH&1&&2&93,040&10WEReSe&\cite{10WEReSe.FeH}&empirical & \ai\\
TiH&1&&3&181,080&05BuDuBa&\cite{05BuDuBa.TiH}&empirical & \ai\\
CO$_2$&13&1000&1&149,587,373&Ames-2016&\cite{17HuScFr.CO2}&empirical & \ai\\
TiO&1&&13&45,000,000&Schwenke&\cite{98Scxxxx.TiO}&tuned \ai & \ai \\
C$_2$H$_2$&1&1000&1&33,890,981&ASD-1000&\cite{17LyPe.C2H2}&effect. Hamilt. $^c$ & effect. dipole $^c$\\
CrH&1&&2&13824&02BuRaBe&\cite{02BuRaBe.CrH}&empirical & \ai\\
CH$_3$F&1& $400$ &1&1,391,882,159&OYKYT&\cite{18OwYaKu.CH3F}&\ai & \ai\\
\bottomrule
\end{tabular}}
\begin{tabular}{@{}c@{}}
\multicolumn{1}{p{\textwidth -.88in}}{\footnotesize $N_{\rm iso}$: Number of~isotopologues considered;
$T_{\rm max}$: Maximum temperature for which the line list is complete (where stated);
$N_{\rm elec}$: Number of electronic states considered;
$N_{\rm lines}$:  Number of lines: value is for the main isotopologue;
DSName: Name of line list and of data set in ExoMol database~\cite{jt631};
$^a$ ExoMol methodology, see text;
$^b$ A mixture of empirical and \ai\ values of the dipole moment function;
$^c$ Empirical, based on Effective Hamiltonian and Dipole moment expansions. }
\end{tabular}
\end{table}
\unskip

{\bf H$_2$O}: the spectrum of water (see Figure~\ref{f:water})
is important in a whole range of astronomical objects
\cite{09Bexxxx.exo} including M-dwarfs~\cite{jt143} and exoplanets
\cite{jt521}. The~ExoMol line lists for H$_2$$^{18}$O and
H$_2$$^{17}$O complement the H$_2$$^{16}$O BT2 line list~\cite{jt378},
which has been widely used for astronomical studies including, for
instance, as the foundation of the BT-Settl cool star model atmosphere
\cite{BT-Settl}.  A comprehensive line list for HDO is also available
\cite{jt469}.  In fact, we have recently updated BT2 with a
new H$_2$$^{16}$O line list known as POKAZATEL~\cite{jt734}. Besides greatly
improving the accuracy of the previous line list, it also extends the
temperature range beyond 3000 K, which BT2 was designed for.
POKAZATEL has been tested against laboratory spectra recorded in
flames and was found to perform very well~\cite{jt712}.  On the scale of
the figure, BT2 and POKAZATEL are very similar for temperatures below
2000~K, although the detailed line positions given by POKAZATEL are
considerably more accurate (see the laboratory study by Campargue~et~al.~\cite{17CaMiVa.H2O}, for example). However, at
high temperatures, POKAZATEL is much more complete. What is really
noticeable is how flat the water opacity becomes at high temperatures.

Figure~\ref{f:polyhydrides} shows cross sections for a number of other hydrides: methane, silane, phosphine, ammonia and ethylene.

{\bf CH$_4$}: Figure~\ref{f:polyhydrides} shows the methane cross
sections generated by the 10to10 line list, which contains almost
$10^{10}$ transitions.  Other line lists available for methane
are notably the TheoReTS line list(s)~\mbox{\cite{14ReNiTy.CH4,17ReNiTyi}},
and a comprehensive experimental line list from
Hargreaves~et~al.~\cite{12HaBeMi.CH4,15HaBeBa.CH4}. The~10to10
line list has been used to identify spectroscopic features in T-dwarfs
\cite{jt596}. Model calculations have shown that reproducing the
atmospheric opacity of these methane-rich brown dwarfs requires the
explicit consideration of many billions of transitions~\cite{jt572}.
As the temperature range of 10to10 is limited, we have recently
extended it with a new 34to10 line list, which contains $34 \times
10^{10}$ transitions~\cite{jt698}. To~make use of this line list
tractable in radiative transport models, the~weak lines are
amalgamated into so-called super-lines~\cite{TheoReTS}.
The~temperature coverage for the various methane line lists, which broadly
agree with each other, is probably now adequate for astrophysical purposes.
However, there is a need for better coverage at shorter wavelengths (<$0.85$~$\upmu$m).


{\bf PH$_3$}:  the phosphine cross sections (see Figure~\ref{f:polyhydrides})
show a very pronounced feature about 4.5 $\upmu$m, which displays only weak
dependence on temperature. Phosphine is a well-known component of solar
system gas giants~\cite{09FlOrTe.PH3}, and it can be anticipated that
this feature will at some point be used to identify PH$_3$ in exoplanets.

{\bf NH$_3$}: the spectrum of ammonia is given in
Figure~\ref{f:polyhydrides}.  Ammonia spectra are well known in brown
dwarfs~\cite{jt596} and are thought to provide the key signature for
coolest class of these species known as Y-dwarfs. Plotted is the BYTe
line list, which is significantly less accurate than most of the other
line lists considered here and is really only complete for wavenumbers
below 10,000 \cm\
 and $T < 1500$~K.  A new ExoMol line list that should help resolve
these issues is currently under construction. A line list for $^{15}$NH$_3$
is also available~\cite{15Yurche.NH3}.

{\bf C$_2$H$_4$}: the final spectrum given in
Figure~\ref{f:polyhydrides} is of ethylene. Only one temperature is
plotted since, despite containing 60 billions line,
 the line list is only complete up to 700 K. However, it is
likely that ethylene will decompose at higher temperatures. Although
not shown here, the~main features of the ethylene spectrum show an
unusually small sensitivity to temperature~\cite{jt729}. Rey~et~al.
\cite{16ReDeNi} have also computed a far-infrared  ethylene line list.

Figure~\ref{f:polyoxides} shows cross sections for a variety of polyatomic oxides plus
hydrogen cyanide.

{\bf H$_2$O$_2$}: hydrogen peroxide cross sections are given
in Figure~\ref{f:polyoxides}. It should be noted that even at room temperature
the line list for H$_2$O$_2$ given by the current release of HITRAN~\cite{jt691} is not complete, as the strong mid-infrared absorptions are missing.
This~is because of the difficulty of analyzing the observed spectra in this
region.

 {\bf HCN}: the line list shown in Figure~\ref{f:polyoxides} is
unusual for two reasons. First, it contains transition sets for two
isomers, HCN and HNC, which are often considered to be  distinct molecules
and second it is not new; rather it is a reworking of earlier line lists
\cite{jt298,jt374} based on extensive compilations of empirical energy
levels due to Mellau~\cite{11Mexxxx.HNC,11Mexxxx.HCN,11Mexxxc.HCN}. A
similar empirically-informed line list is available for H$^{13}$CN
\cite{jt447}. The~updated Harris line list was used to tentatively
identify HCN in super-Earth exoplanet 55 Cancri e~\cite{jt629}.
However, the 1.54~$\upmu$m feature used for this detection is close to a
similar acetylene feature. As discussed below, there is still no good
line list for hot acetylene.  The~opacity of HCN is particularly
important for reliable models of cool carbon stars where its inclusion
leads to fundamental changes in the stellar structure
\cite{84ErGuJo.HCN}.  The~ratio between HCN and HNC can provide a
potential thermometer the atmospheres of cool stars~\cite{jt304}.

{\bf SO$_2$}: sulphur dioxide cross sections are given in
Figure~\ref{f:polyoxides}. SO$_2$ is thought to be an important component of
oxygen-rich exoplanet atmospheres~\cite{10KaHeSa.SO2,13HuSeBa.SO2}
and to have been important in both early-Earth~\cite{13WhXiHu.SO2} and
early-Mars~\cite{09JoPaMi.SO2}.

{\bf SO$_3$}: sulphur trioxide cross sections are given in
Figure~\ref{f:polyoxides}. SO$_3$ can be hard to detect because it usually
appears in conjunction with SO$_2$, which absorbs strongly in the same region.
Interestingly, our cross sections suggest that it may be easier to distinguish
SO$_3$ from SO$_2$ at higher temperatures as, while the spectrum
SO$_3$ retains much
of it structure at higher temperatures, the~spectrum of SO$_2$ becomes
increasingly flattened. The~line list is only complete up to 800~K
and this has been used as the upper temperature.

{\bf H$_2$CO}:  Figure~\ref{f:polyoxides} shows temperature dependent
cross sections of formaldehyde.

{\bf CO$_2$}: carbon dioxide cross sections given in
Figure~\ref{f:polyoxides} are obtained from the Ames-2016 (4000~K) line
list of  Huang~et~al.~\cite{14HuGaFr.CO2,17HuScFr.CO2},
for the main isotopologue ($^{12}$C$^{16}$O$_2$) and covers rotational
excitations up to $J$ = 220.  The~line list is based on an empirical
potential energy surface and accurate variational nuclear motion
calculations. The~Ames-2016 database also contains line lists for
12 other~isotopologues of CO$_2$.

Figure \ref{f:HNO3:CH3Cl:C2H2} shows cross sections of HNO$_3$, CH$_3$Cl and C$_2$H$_2$.

{\bf HNO$_3$}: nitric acid, whose spectrum is shown in
Figure~\ref{f:HNO3:CH3Cl:C2H2}, is not likely to be an important source of opacity.
However, its spectroscopic signature
can be clearly observed in the Earth's atmosphere from space
\citep{96BlDe.HNO3,11CoMaSa.HNO3,18ScStHe} and thus HNO$_3$ could be a possible
biomarker on exoplanets. Like H$_2$O$_2$, the~HITRAN data for  HNO$_3$ is
incomplete for frequencies above  about 1000 \cm.

{\bf CH$_3$Cl}: methyl chloride is also a potential
biosignature in the search for life outside of the solar system
\cite{13SeBaHu.exoplanet}. ExoMol's mew line list for methyl
chloride covers the two major isomers, CH$_3^{35}$Cl and~CH$_3^{37}$Cl.

\textls[-15]{{\bf C$_2$H$_2$}: acetylene cross sections given in
Figure~\ref{f:HNO3:CH3Cl:C2H2} are generated using the ASD-1000
database~\cite{17LyPe.C2H2}. ASD-1000 was generated
using an Effective Hamiltonian and contains about 34,000,000 lines. This~would appear to be too few lines for the line list to be complete
at  elevated temperatures as
normally, hot line lists for tetratomics contain billions of
transitions in order to be complete for high temperatures. A variational
acetylene line list is currently under construction as part of the
ExoMol project.}

{\bf H$_2$S}: Figure~\ref{f:H2S} shows that hydrogen sulphide
has a rather unusual spectrum. Transitions involving the vibrational fundamentals are anomalously weak~\cite{98BrCrCr.H2S} for this system, which shifts the peak of the
vibration--rotation spectrum to shorter wavelengths than is usual: the
figure shows that the strongest vibration band is an overtone lying in the
2.5 $\upmu$m region.

Figure~\ref{f:XH}  gives cross sections for NS as well as  the  alkaline earth monohydrides MgH and CaH. Together with BeH, MgH and CaH  were the first set of species considered by ExoMol. These~ ExoMol line lists only contain pure rotation and rotation--vibration transitions within the X~$^2\Sigma^+$ electronic ground states for each of these
species.

\unskip

{\bf NS}: NS is one of the first ten diatomic molecules to be
detected in space \citep{77Somerv.NS, 79LoJoSn.NS}.

{\bf MgH}: the A~$^2\Pi$--X~$^2\Sigma^+$ band is also
important for magnesium monohydride as, in particular, its use of
major importance for establishing isotopic abundances of Mg in stars
\cite{80ToLaxx.MgH,
  88McLaxx.MgH,00GaLaxx.MgH,03DaYDaL.MgH,13HiWaRa.MgH}.
Weck~et~al. ~\cite{03WeScSt.MgHline} constructed a
purely \ai\ line list, which covers this band but is of
limited accuracy. More recently, experimentally-driven studies
\cite{13GhShBe.MgH,13HeShTa.MgH} have sought to remedy this problem.

{\bf CaH}: A~$^2\Pi$--X~$^2\Sigma^+$ band is also important in cool stars and brown dwarfs~\cite{07ReHoHa.CaH}.

Figure~\ref{f:BeH:AlH:CH} shows cross sections covering electronic spectra of three diatomic monohydrides,  BeH, AlH and CH.

{\bf BeH}:  Darby-Lewis~et~al.~\cite{jt722} have recently developed a new model for beryllium monohydride,
which both improves (marginally) on the earlier Yadin line list~\cite{jt529} and,
more significantly, includes a treatment of the A~$^2\Pi$--X~$^2\Sigma^+$
band. This~band has been observed in sunspot~\cite{08ShBaRa.BeH}, but
the motivation for extending the line list was actually fusion plasmas
where the use of Be walls has led to the observation of A--X emissions
in the plasma~\cite{98DuStSu.BeH}.
Darby-Lewis~et~al. demonstrate that their new line list gives excellent reproduction of emission spectra of both BeH and BeD, as well as predictions for the spectrum of BeT, which is important for fusion studies.

{\bf AlH}: it is known to be present in sunspots through lines in its A~$^1\Pi$--X~$^1\Sigma^+$ electronic
band which lie in the blue \citep{00WaHiLi}; AlH was also recently detected
around Mira-variable o Ceti by Kaminski~et~al.~\cite{16KaWoSc.AlH}. The~ExoMol line list for AlH contains these electronic transitions as well as  ground state lines. 

\textls[-15]{{\bf CH}: an extensive, empirical  line list is available
 for CH due to Masseron~et~al.~\cite{14MaPlVa.CH}. This~line list covers transitions
within the X~$^2\Pi$ electronic ground state as well as rovibronic
transitions within the A~$^2\Delta$--X~$^2\Pi$,
B~$^2\Sigma^-$--X~$^2\Pi$ and C~$^2\Sigma^+$--X~$^2\Pi$ bands.
 Masseron~et~al.~\cite{14MaPlVa.CH} also provide a line list for $^{13}$CH.}

Figure~\ref{f:HCl:HS:CrH:OH} shows temperature dependent cross sections for HCl, HS, CrH, and OH.

\textls[-20]{{\bf HCl}: the line list used to generate the HCl cross sections
given in  Figure~\ref{f:HCl:HS:CrH:OH} was generated alongside line lists
for other hydrogen halides, namely HF, HBr and HI, using semi-empirical
methods by Li.~et~al.~\cite{11LiGoBe.HCl,13LiGoHa.HCl}.}

Figure~\ref{f:polar} shows cross sections for NaCl, KCl, PO, PN and CS.

{\bf NaCl} and {\bf KCl}: cross sections are shown in Figure~\ref{f:polar}. These species have large transition dipoles
and hence strong transitions, and they are thought to be of importance for
studies of hot super-Earths~\cite{jt693}.

{\bf PO}: phosphorus monoxide has been detected in a number of locations in space: red Supergiant Star VY Canis Majoris~\cite{07TeWoZi.PO},
in the wind of the oxygen-rich AGB star IK Tauri~\cite{13DeKaPa.PO}, and in star-forming
regions~\cite{16RiFoBe.PO,16LeVaVi.PO}. The~ExoMol line list covers transitions within the  X~$^2{}\Pi$ ground electronic state.

{\bf PN}: it  is an  important molecule used to probe different regions of the interstellar medium (ISM). The~ExoMol line list for PN covers just pure rotation--vibration transitions.

{\bf CS}: cross sections in Figure~\ref{f:polar} illustrate the temperature dependence of the CS spectra as generated using the
comprehensive vibration--rotation ExoMol line lists for eight~isotopologues
of carbon monosulphide (CS) ($^{12}$C$^{32}$S, $^{12}$C$^{33}$S,
$^{12}$C$^{34}$S, $^{12}$C$^{36}$S, $^{13}$C$^{32}$S,
$^{13}$C$^{33}$S, $^{13}$C$^{34}$S, $^{13}$C$^{36}$S) in their
ground electronic states.

Figure~\ref{f:cdiatomics} shows cross sections for the important species
CO plus CN, CP and CaO.

{\bf CO}: cross sections for carbon monoxide are given in
Figure~\ref{f:cdiatomics}. CO is ubiquitous throughout the spectra
of cool stars~\cite{jt352} and brown dwarfs,
and can clearly be seen in exoplanets~\cite{13DeBrSn.exo,14BrDeBi.exo}.
CO is a very strongly bound species so the molecule survives at relatively
high temperatures. As can be seen from the figure, its broadband spectrum
changes from comprising a series of sharp bands to becoming quasi-continuous
as the temperature is raised.
Unlike the other species shown in this figure, the~CO cross sections
through the visible are simply given by rotation--vibration transitions. We note that
Li.~et~al.~\cite{15LiGoRo.CO} improved their
intensities by refining  their CO \ai\ dipole moment curve by fitting  to experiment.
This~is not the usual practice~\cite{jt573}, but Li.~et~al. found that
they could not get satisfactory results using a purely \ai\ approach.

{\bf CN} and {\bf CP}: cross sections are shown in
Figure~\ref{f:cdiatomics}. These two open shell or radical molecules have
the same electronic structure and their spectra are dominated by the A
$^2\Pi$--X $^2 \Sigma^+$ and B $^2\Sigma^+$--X $^2\Sigma^+$ bands, which
for CN lie in the red and violet regions of the visible. For the
heavier CP radical, these bands are shifted to longer wavelengths.

{\bf CaO}: cross sections for calcium oxide are given in
Figure~\ref{f:cdiatomics}. CaO is unlikely to exist in the gas phase
in dense atmospheres at 300 K, this spectrum is shown for the consistency.
 CaO has yet to be detected in space, but it is thought of as a possible
component of super-Earth exoplanet atmospheres. CaO possesses transitions
with notably large cross sections, which should facilitate its detection.

Figure~\ref{f:NO} shows cross sections for NO and PS.

{\bf NO} and {\bf PS}: are both radicals with X~$^2\Pi$
electronic ground states. NO is an important species in the Earth's
atmosphere and is likely to be so in oxygen-rich exoplanets. The
current ExoMol line list~\cite{jt686} supersedes the one provided by
HITEMP~\cite{jt480}.  For NO, the transitions considered in the figure
are all within the  X~$^2\Pi$
ground state manifold, while, for PS, the B~$^2\Pi$--X~$^2\Pi$
electronic band is also considered. The~equivalent electronic
transition for NO, which is actually designated the A~$^2\Pi$ --
X~$^2\Pi$ band, lies well into the ultra violet at about 230 nm
\cite{97DaDoKe.NO}.

Figure~\ref{f:ZH} shows cross sections for
LiH, ScH, FeH and NH.

\textls[-15]{{\bf LiH} cross sections are shown in Figure~\ref{f:ZH}.
They~are based on \ai\ calculations by \mbox{Coppola~et~al. }~\cite{jt506}. LiH is a four electron system so purely
theoretical procedures should give reliable results. The~spectrum of LiH
is important for possible applications of the lithium test in
brown dwarfs~\cite{97Pavlen}. We note that very recently
Bittner and Bernath~\cite{18BiBexx} have constructed line lists
for both LiF and LiCl.}

\textls[-15]{{\bf ScH} cross sections are shown in Figure~\ref{f:ZH}. These were also based on an essentially \ai\ line list, which
contains an empirical adjustment for the band positions. Unlike LiH,
ScH is a many-electron system with an open $d$-shell; such systems
present considerable challenges from an \ai\ perspective~\cite{jt623,jt632}; the provision of high accuracy line list for ScH
will probably require further experimental input.}

\textls[-15]{{\bf FeH} cross sections, shown in Figure~\ref{f:ZH}, are
conversely based on experimental studies. FeH is an important
stellar species, which has been observed in M and L dwarfs~\mbox{\cite{03CuRaDa.FeH,10WEReSe.FeH,10HaHiBa.FeH}} as well as sunspots~\mbox{\cite{01WalHin.FeH,04RaBuCo.FeH,08HarBro.FeH}}.  There is a strong need
to use these laboratory studies to construct a rigorous
theoretical model for the system, which can be used to generate
line lists over an extended range of temperatures and wavelengths;
however, \ai\ calculations on the system remain
a challenge~\cite{12DeWexx.FeH}.}

{\bf NH}: an empirical line list for NH from the Bernath group
\cite{14BrBeWe.NH}  covering  rotation--vibrational transitions within the 
X~$^3{}\Sigma^{-}$ ground electronic state.

Figure~\ref{f:YH} shows cross sections for SiH, NaH and TiH.

{\bf SiH}: is an accurate line list covering rotation--vibration transitions within the ground X~$^{2}\Pi$ electronic state as well as transitions to the low-lying A~$^{2}\Delta$  and a~$^{4}\Sigma^-$ states.

{\bf NaH} cross sections are shown in Figure~\ref{f:YH}.  NaH
remains undetected in stellar spectra but is thought to be an
important opacity source in M-dwarfs~\cite{13RaReAl.NaHAlH}. However,
as our cross sections show the blue A $^2\Sigma^+$--X $^2\Sigma^+$
band is extended and gives a rather flat band shape. This~will make
the clear detection of NaH in a stellar atmosphere difficult from its
main electronic band.

\textls[-0]{{\bf TiH} cross sections shown in Figure~\ref{f:YH} represent an empirical line list for TiH from \mbox{Burrows~et~al.}~\cite{05BuDuBa.TiH}.}

Figure~\ref{f:YH} shows cross sections for SiO, AlO, VO and TiO.

{\bf SiO} cross sections are shown in Figure~\ref{f:XO}. SiO
is a strongly bound molecule whose infrared spectrum is well-known
in stars~\cite{10ArVaMa.SiO} and sunspots~\cite{95CaKlDu.SiO,02WaHixx.SiO}.
Its electronic spectrum has also been observed in sunspots~\cite{79JoPuPa.SiO}. At present, the ExoMol project has only provided a long-wavelength
line list for SiO based on its temperature-dependent vibration--rotation and
rotation spectrum.  Work is in progress providing a companion line list
covering its electronic bands starting from \ai\ potential energy curves recently calculated by Bauschlicher Jr.~et~al.~\cite{16Bausch.SiO}. In order to illustrate the UV spectra of SiO, Figure~\ref{f:XO} also shows cross sections from Kurucz'z database~\cite{11Kurucz.db}.

{\bf AlO}: the spectrum of aluminium monoxide is illustrated
in  Figure~\ref{f:XO}. This~spectrum is actually important
for plasma diagnostics~\cite{14BaMoLe.AlO,14SuPa.AlO}. Rather remarkably,
the only previous AlO line list~\cite{11PaHoxx.AlO}, which was
prepared for plasma studies, only provided relative as opposed to
absolute line intensities.

{\bf VO} cross sections are shown in Figure~\ref{f:XO}. The
VO spectrum is heavily overlapped by that of TiO and the two molecules often
occur together. VO spectral signatures are
important for classifying  late M dwarfs ~\cite{93KiKeRi.VO,04McKiMc.dwarfs},
and it is one of the dominant species in the spectra of young hot brown dwarfs~\cite{04McKiMc.dwarfs,06KiBaBu.dwarfs,08PeMeLu.dwarfs}.
VO is thought to be present in the atmosphere
of hot Jupiter exoplanets~\cite{08DeVide.TiO}.

{\bf TiO} cross sections are shown in Figure~\ref{f:XO}. TiO is
a well-known and important molecule in the atmosphere of M-dwarf stars
\cite{00AlHaS1.TiO}. Detections in the atmospheres of exoplanets are
also beginning to be reported~\cite{17SeBoMa.TiO,17NuKaHa.TiO,jt699}.
The~figure is based on the
line list of Schwenke~et~al.~\cite{98Scxxxx.TiO}; there is also
a line
list due to Plez~\cite{98Plxxxx.TiO}, which
has been updated in a recent release of the VALD database~\cite{VALD3}.
However, there are known problems with the available spectroscopic
data on TiO (see Hoeijmakers~et~al.~\cite{15HoDeSn.TiO}, for example).
An ExoMol TiO line list is currently nearing completion~\cite{jtTiO}; this
line list will make use of the MARVEL study on TiO~\cite{jt672}, which
leads to more accurate transition frequencies than are provided by
the currently available line lists.

Figure~\ref{f:uv} gives cross sections for the open shell diatomic species
C$_2$ and of the important molecular ion H$_3^+$.

{\bf C$_2$} has a long spectroscopic history
\cite{1802Wo.C2,1857Swan.C2}. In astronomical
objects, its spectrum  has been observed via at least six
distinct electronic bands, with other band systems required to explain
observed populations. An empirical line list for the Swan system was
provided by Brooke~et~al.~\cite{13BrBeScBa.C2}; the figure
shows cross sections generated using the newly constructed ExoMol line
list~\cite{jtexoC2}, which considers transitions between the eight lowest
electronic states in the system. Given the many perturbations between
the electronic states that are hard to model and that there are astronomically
observed bands not covered by this eight-state model, further work on C$_2$
will undoubtedly be needed.


{\bf H$_3^+$}: the spectrum of  H$_3^+$ given in
Figure~\ref{f:uv} is based on \ai\ calculations,
which have proved themselves to be very accurate for this molecular
ion~\cite{jt512}.  H$_3^+$ is an unusual species in that it has
no (allowed~\cite{jt72}) rotational spectrum and no known electronic
spectrum leaving only rotation--vibration transitions. These~ are
well-known in the ionospheres of  solar system gas giant planets~\cite{jt258}
but have yet to be observed in hot Jupiter exoplanets~\cite{06ShGaMo.H3+},
where they might be thought to be prominent. However, the~cooling provided
by H$_3^+$ infrared emissions appears to be the key for the
stability of atmospheres of giant extrasolar planets close to their host star
\cite{kam07}. This~has made H$_3^+$ cooling functions of great importance
\cite{jt489,jt551}. Similarly H$_3^+$ has yet to observed in stellar
or brown dwarf spectra, but H$_3^+$ is the main source of electrons in
 objects such as cool white dwarfs~\cite{97BeRuLe.H3+}. The~partition
function of H$_3^+$~\cite{jt169} thus controls the amount of H$^-$
present, which, in turn, dominates the opacity.
A~line list for H$_2$D$^+$~\cite{jt478} is also available.

\section{Conclusions}

Compiling molecular opacity functions requires a large range of
spectroscopic data on a large range of molecules. As discussed in this
article, for many key species, there are now extensive line lists available
that can be used to compute temperature-dependent opacity
functions. This~is a process of constant improvement and immediate improvements
are indeed identified for several species discussed
above. Here, we consider what species may be missing from the current
compilations.

\textls[-25]{In their validation of the well-used BT--Settl model~\cite{BT-Settl}
atmospheres for M-dwarfs, Rajpurohit~et~al.~\cite{13RaReAl.NaHAlH} identified only AlH,
NaH and CaOH as key species for which data was missing. In response to
this, the ExoMol project has provided spectroscopic line lists for AlH
\cite{jt732} and NaH~\cite{jt605}; this leaves CaOH, which has a strong
band in the visible at 557 nm, as the one key missing species.}

The~chemistry of oxygen-rich M-dwarf stars is somewhat simpler than that
of carbon-rich stars. There are a number of carbon-containing species for
which reliable line lists are not available, notably acetylene (HCCH)
and C$_3$. Acetylene spectra are well-studied
experimentally~\cite{jt705,03HeCaEl.C2H2,07Hexxxx.C2H2} and recently
Lyulin and Perevalov~\cite{17LyPe.C2H2} produced an effective-Hamiltonian
based, empirical $^{12}$C$_2$H$_2$ line list. Work~on an ExoMol acetylene
line list is nearing completion. J{\o}rgensen~et~al.~\cite{89JoAlSi.C3}
produced an early, purely \ai\ line list for C$_3$. They demonstrated the importance of this species for models of cool carbon stars, but their
line list cannot be considered reliable by modern standards where \ai\ methods
have improved considerably and tuning models to experimental data is now routine. The~construction
of an improved C$_3$ line list is underway as part of the ExoMol project.
Both ExoMol and TheoReTS are considering line lists for higher hydrocarbons
that are needed for studies of~exoplanets.

Transition metal containing diatomics provide strong sources of opacities
since these open shell species often display strong electronic transitions
in the near infrared and visible i.e., near the peak of the stellar
flux for a cool star. A few of these species, notably TiO, VO, FeH, ScH and
TiH, are considered above but there are many more possible candidates.
These~ include CrH~\cite{jtCrH}, MnH~\cite{jtMnH} and MgO ~\cite{jtMgO} for which ExoMol has near complete line lists, and species such  NiH, ZrO, YO, FeO. Carbides,
nitrides and even sulfide species may also need to be considered in due
course. Performing accurate \ai\ electronic structure
calculation on transition metal containing molecules remains very difficult
\cite{jt623,jt632}.

Recent observations have identified a whole new class of exoplanets
with masses somewhat larger than the Earth's and orbits close to their
host stars. These~ hot rocky super-Earths, lava planets or magma
planets as they are variously known as are just beginning to be
characterized~\cite{jt629}.  Even~though at this point the molecular
composition of their atmosphere is essentially unknown, study of these
bodies will create demands for spectroscopic data and opacities for a
whole range of new molecules.  We recently reviewed the data needs for
hot super-Earth exoplanets~\cite{jt693} and refer any interested
reader to that article.

Finally, line broadening must be mentioned. Molecular transitions
undergo broadening due to temperature and pressure effects. While
temperature-induced Doppler broadening is straightforward to model,
pressure broadening is more difficult because its importance is, at least
in principle, different for every transition. Broadening parameters
for key collision broadeners such as H$_2$ and He at elevated temperatures
are largely missing. Approximate methods are being used to fill this
void for key molecules such as water~\cite{jt544,jt699,18ReClHa}. We have
implemented a systematic and complete if approximate broadening
``diet''~\cite{jt684} as part of the ExoMol data base~\cite{jt631}.
This~allows the generation of both temperature- and pressure-dependent
cross sections using our program ExoCross~\cite{jt708}. However, there
remains much work to be done in both improving  temperature-dependent
models for line broadening and providing the appropriate broadening
parameters.
\vspace{6pt}

\acknowledgments{We thank the  members of the ExoMol team
for their work on this project. The~ExoMol project was funded
by the ERC Advanced Investigator Project 267219 and is currently
supported by UK STFC under grant ST/M001334/1.}

\authorcontributions{Both authors contributed equally.}


\conflictsofinterest{The~authors declare no conflict of interest.}


%


\bibliographystyle{mdpi}
\reftitle{References}

\end{document}